\documentclass
[10pt,superscriptaddress,secnumarabic,amssymb,amsmath,nobibnotes,aps,prd,nofootinbib]{revtex4}%
\usepackage{url}
\usepackage[T1]{fontenc} % if needed
\usepackage[utf8]{inputenc}
\usepackage{mathtools}
\usepackage{graphicx}
\usepackage{multirow}
\usepackage{enumerate}
\usepackage{float}
\usepackage{textcomp}
\begin{document}
	
\markboth{A. Nájera and A. Fajardo}
{Fitting $f(Q,T)$ gravity models with a $\Lambda$CDM limit using H(z) and Pantheon data}

\title{Fitting $f(Q,T)$ gravity models with a $\Lambda$CDM limit using H(z) and Pantheon data}  
\author{ANTONIO NÁJERA}

\address{Instituto de Ciencias Nucleares, Universidad Nacional
  Aut\'onoma de M\'exico, Circuito Exterior C.U., A.P. 70-543 \\
  M\'exico D.F. 04510, M\'exico\\
antonio.najera@correo.nucleares.unam.mx}

\author{AMANDA FAJARDO}

\address{Facultad de Ciencias Físico-Matemáticas, Universidad Autónoma de Nuevo León, Av. Pedro de Alba S/N, Ciudad Universitaria\\
Monterrey, Nuevo León 66455, México\\
amanda.fajardogr@uanl.edu.mx \\
\sffamily\textcopyright 2021. This manuscript version is made available under the CC-BY-NC-ND 4.0}

\begin{abstract}
	We proposed five $f(Q,T)$ models, which are an extension of symmetric teleparallel gravity, where $Q$ is the non-metricity and $T$ is the trace of the stress-energy tensor. By taking specific values of their parameters, these models have a $\Lambda$CDM limit. 
	Using cosmic chronometers and supernovae Ia data, we found that our models are consistent with $\Lambda$CDM at a 95\% confidence level. To see whether one of these models can challenge $\Lambda$CDM at a background perspective, we computed the Bayesian evidence for them and $\Lambda$CDM. According to it, the concordance model is preferred over four of them, showing a weak preference against $f(Q,T) = -Q/G_N + bT$ and $f(Q,T) = -(Q+2\Lambda)/G_N +bT$, a substantial preference against $f(Q,T) = -(Q+2 H_0^2 c (Q/(6H_0^2))^{n+1})/G_N + bT $, and a strong preference against $f(Q,T) = -(Q+2H_0^2c(Q/(6H_0^2))^{n+1} + 2\Lambda)/G_N + bT$. Interestingly, a model includying a $T^2$ dependence ($f(Q,T) = -(Q+2\Lambda)/G_N - ((16\pi)^2 G_N b)/(120 H_0^2) T^2$) showed a {substantial} preference against $\Lambda$CDM. {Therefore, we encourage further analyses of this model to test its viability outside the background perspective.}
\end{abstract}


\maketitle

\section{Introduction}
\label{sec:introduction}
The current standard model of cosmology $\Lambda$CDM uses general relativity (GR) and a cosmological constant $\Lambda$ that acts as an "energy of vacuum" or "dark energy". This dark energy fills the space-time and drives the present-day cosmic acceleration \cite{peebles2003cosmological}. Although the model is successful on cosmological scales, {it has some theoretical problems. Ror example, the value of the observational cosmological constant is in disagreement with its theoretical prediction from quantum field theory} \cite{Bull_2016}. Moreover, { it has observational problems such as the determination of the Hubble constant because the Planck and SH0ES collaborations give estimations that are in more than $4\sigma$ tension between each other.}  \cite{Riess_2019,aghanim2020planck}.\\

{There are many attempts to modify Einstein's original equations to account for cosmological expansion. One of them are modified gravity theories, such as $f(R)$ gravity, $f(T)$ gravity and modified Gauss-Bonnet gravity, to name some.} A different geometrical description of gravity comes from symmetric teleparallel gravity. Here, the curvature and the torsion tensor vanishes. However, the non-metricity, defined as the divergence of the metric, is different from zero and carries the gravitational force. Einstein used its predecessor, teleparallel gravity, in a unification theory attempt and GR can be written in terms of this geometry \cite{nester1998symmetric}. This theory shows great advantages. For example: a tensorial Hamiltonian that has a positive energy proof, a stress-energy true tensor, a translational gauge nature, and —in a Minkowski space-time— all gravitational waves propagate at the speed of light \cite{xu2019f,arora2021constraining,nester1998symmetric,arora2020f}.  \\

A novel $f(Q, T)$ theory, extension of symmetric teleparallel gravity, was recently proposed by Xu et. al. \cite{xu2019f}. In this theory, $Q$ is the non-metricity and $T$ the trace of the stress-energy tensor. This model shows accelerating and decelerating phases and its baryon-to-entropy ratio is very close to the observational value \cite{arora2020f,Bhattacharjee_2020}. Moreover, it is {consistent with the data from the Baryon Acoustic Oscillations (BAO), the Supernovae Ia (SNeIa) and the Differential Ages Method (DA)} (which uses estimations from differential ages of galaxies). A deeper analysis in $f(Q,T)$ theory and some of it applications can be found in \cite{arora2020f,arora2021constraining,Arora_2020energ,Arora_2021}\\

One of the recent successes of this novel theory is the model that Xu et. al. proposed, which was tested in \cite{arora2020f} with BAO and SNeIa data. They found values of the acceleration and deceleration parameters for the model $f(Q,T)=mQ^n+bT$ (with $m$,$n$ and $b$ as model parameters) that are consistent with the present accelerating universe. They also found that $f(Q,T)$ models can be a geometrical description of dark energy.\\

At a perturbative level, the $f(Q,T)$ models have not been studied yet. Nevertheless, the $\Lambda$CDM sub-model is an $f(Q)$ function only as we will show in section \ref{sec:models}. 
An $f(Q)$ model of the form $f(Q)=Q+M\sqrt{Q}+2\Lambda$ (where M is a free parameter, $\Lambda$ the cosmological constant and Q the non-metricity) is a better fit for the redshift space distortions data than $\Lambda$CDM, and relieves the tension of the amplitude of density fluctuations between the Planck nd Large Scale Structure data\cite{Barros_2020}. Moreover, in \cite{khyllep2021cosmological} was shown that for a model of the form $f(Q)=Q+\alpha Q^n$ (with $\alpha$ and n as model parameters) the asymptotic value of the growth index depends solely on the parameter $n$, and for $n<0$ this value is smaller than the one from $\Lambda$CDM. The models presented here have one of these forms when b (the parameter accounting for deviations from pure $f(Q)$ theory) is zero.\\

{Since we want our models to have a $\Lambda$CDM limit, we build $f(Q,T)$ functions that have $\Lambda$CDM as a sub-model. However, a further analysis must be done to test if these models have stable perturbations and satisfy all the theoretical and observational requirements of a good cosmology model}. We can recover {$\Lambda$CDM as a sub-model in two ways. The first one is} to use $f(Q)=-\dfrac{1}{G_N}(Q+cQ^{1/2}+6\Lambda)$ because it reproduces the $\Lambda$CDM Friedmann equations \cite{ayuso2021observational} (with $c$ as a parameter). If we generalize this model into the form $f(Q,T)=-\dfrac{1}{G_N}(Q+6\Lambda)+bT$, we are able to account for deviations from the standard model via the parameter $b$. The second form is to use functions of the form $f(Q,T)=-\dfrac{Q}{G_N}+bT$ and take a parametric ansatz of the effective equation of state $w=-\dfrac{1}{1+m(1+z)^n}$ to replicate the Hubble function of $\Lambda$CDM with $b=0$ and $n=3$ \cite{arora2021constraining}. The first form includes the cosmological constant to account for the cosmological expansion while in the second form the acceleration and deceleration phases are a result of the EoS chosen. \\

{All the models presented here are extensions of either one of these two forms.} {In this paper, we considered the {effective} Equation of State (EoS) $w$ representing all energy components of the Universe together.} \\

In three models, we assumed that the Universe is made up of only dust pressure-less matter plus a cosmological constant. These are \ref{model:3} $f(Q,T) = -(Q + 2\Lambda)/Q_N + bT$,  \ref{model:5} $f(Q,T) = -(Q + 2H_0^2 c (Q/(6H_0^2))^{n+1} + 2\Lambda)/G_N + bT$, and \ref{model:6} $f(Q,T) = -(Q+2\Lambda)/G_N - ((16\pi)^2 G_N b)/(120 H_0^2) T^2$ . In addition to that, we considered two models whose equation of state is a parametrization of two constants $m ,\, n$. These are \ref{model:2} $f(Q,T) = -Q/G_N + bT$ and  \ref{model:4} $f(Q,T) = -(Q + 2H_0^2 c (Q/(6H_0^2))^{n+1})/G_N + bT$. \\

To constraint the model parameters, we utilized the cosmic chronometers (CC) compilation of \cite{magana2018cardassian}, $H_0$ (the value of the SH0ES collaboration's Hubble constant) and type Ia Supernovae data (SNeIa) of \cite{scolnic2018complete}. To compare the models with $\Lambda$CDM we used the criterion of the Bayes evidence.   \\

This article is organized as follows: in section \ref{sec:f(Q,T)theory}, we give a general overview of the theory. The presentation of the specific models is given in section \ref{sec:models}. In section \ref{sec:obsdatabase}, we show the data sets used and the $\chi^2$ functions to minimize. The statistical analysis using the Bayes evidence criteria is done in sections \ref{sec:methodology} and \ref{sec:model}. Finally, some final remarks are presented in section \ref{sec:Conclusions}.

\section{$f(Q,T)$ Theory}
\label{sec:f(Q,T)theory}

The general action as first proposed in \cite{xu2019f} is
\begin{align}
	S=\int \left(\frac{1}{16\pi}f(Q,T)+\mathcal{L}_{m} \right)\sqrt{-g}d^4 x ,
\end{align}
where $\mathcal{L}_m$ is the matter Lagrangian, Q the non-metricity {and $T$ the trace of the stress-energy tensor}. \\
Using the metric of a homogeneous, isotropic and spatially flat universe
\begin{align}
	ds^2= -dt^2+a^2(t)\delta_{ij}dx^i dx^j ,
\end{align}
the generalized Friedmann equations for this gravity theory are \cite{xu2019f}
\begin{equation}
	\label{eqn:firstFriedmannEquation}
	8 \pi \rho = \frac{f}{2} - 6FH^2 - \frac{2\Tilde{G}}{1+\Tilde{G}} (\dot{F} H + F \dot{H})
\end{equation}
\begin{equation}
	\label{eqn:secondFriedmannEquation}
	8\pi p = -\frac{f}{2} + 6FH^2 + 2(\dot{F} H + F \dot{H}) ,
\end{equation}
where $8\pi\tilde{G}=f_T$ and $F=f_Q$ are the derivatives of $f(Q,T)$ with respect to its variables. Combining both we get the evolution of $\dot{H}$
\begin{equation}
	\label{eqn:evolutionH}
	\dot{H} + \frac{\dot{F}}{F} H = \frac{4\pi}{F} (1+\Tilde{G}) (\rho + p) . 
\end{equation}

We can also write equations (\ref{eqn:firstFriedmannEquation}) and (\ref{eqn:secondFriedmannEquation}) in a similar form to the standard Friedmann equations of GR
\begin{equation}
	\label{eqn:standardFirstFriedmannEquation}
	3H^2 = 8\pi \rho_{\text{eff}} = \frac{f}{4F} - \frac{4\pi}{F} \left[ (1+\Tilde{G})\rho + \Tilde{G} p \right]
\end{equation}
\begin{equation}
	\label{eqn:standardSecondFriedmannEquation}
	2\dot{H} + 3H^2 = - 8 \pi p_{\text{eff}} = \frac{f}{4F} - \frac{2\dot{F}H}{F} + \frac{4\pi}{F} \left[ (1+\Tilde{G}) \rho + (2+\Tilde{G}) p \right] ,
\end{equation}
where $\rho_{\text{eff}}$ is the effective density and $p_{\text{eff}}$ the effective pressure. It is fair to mention that in $f(Q,T)$ theories, the stress-energy tensor is not conserved. Because of this, the energy balance equation is not zero but given by \cite{xu2019f} 
\begin{equation}
	\dot{\rho} + 3H(\rho + p) = \frac{\Tilde{G}}{16 \pi (1+\Tilde{G})(1+2\Tilde{G})} \left[ \dot{S} - \frac{(3\Tilde{G}+2)\dot{\Tilde{G}}}{(1+\Tilde{G})} S + 6HS \right] ,
\end{equation}
with $S = 2(\dot{F} H + F \dot{H})$.

\section{$f(Q,T)$ Models}
\label{sec:models}

We supposed that the content of the universe consists of a perfect fluid with a stress-energy tensor defined as
\begin{equation}
	\label{eqn:stressEnergyTensor}
	T_{\; \nu}^\mu = \text{diag} (-\rho, p, p, p) .
\end{equation}

Instead of considering $H_0$ as a free parameter, we took $h = H_0 / (100 \, \text{km} \, \text{s}^{-1} \, \text{Mpc}^{-1})$ because it is dimensionless. The models considered here reduce to $\Lambda$CDM for an specific value of their parameters. We want this to happen {because it is the current standard model of cosmology. In a future work we will test these models outside of the $\Lambda$CDM wherear in the present work, we focused on testing the models at a background perspective using low-redshift data. \\
	
	First, we need to find the general form of the function $f(Q, T)$ that not only reproduces $\Lambda$CDM Friedmann equations and its density contents but, in addition, satisfies the $\Lambda$CDM stress-energy tensor conservation. Consequently, the $\Lambda$CDM sub-model should have $\Tilde{G} = 0$ and it is therefore a pure $f(Q)$ function. We used the model given by \cite{ayuso2021observational} as the concordance model. \\
	
	The extensions of $\Lambda$CDM that we considered have the form
	\begin{equation}
		f(Q,T) = aQ + bT^m + cQ^{n+1} + d ,
	\end{equation}
	With $a=-1/G_N$, $b, c, d$ parameters and $m, n$ fixed values for the exponents. {We choose this form of $f(Q,T)$ because it allows us to include a cosmological constant with $d = -(2\Lambda)/G_N$, it includes a $T^m$ term that accounts for deviations from $f(Q)$ to $f(Q,T)$, and it considers a $Q^{n+1}$ term that along with the $T^m$ term representing deviations from $\Lambda$CDM}.\\
	
	In this paper, we only studied expressions with $n=-2$ because they give quadratic equations of the Hubble parameter $H(z)$. Therefore, {the solution gives two branches of $H(z)$. As we will see, the positive branch predicts a late-time accelerated expansion while the negative one predicts a decelerating universe that is inconsistent with observational data}. We will also consider only $m=1,2$ because these exponents of $T$ give linear and quadratic equations of $\rho$ and $p$ \cite{ayuso2021observational}. Different values of $n$ and $m$ would require a careful algebraic treatment to solve third or higher-order equations of $H(z)$ and $\rho, p$. Thus, the branch of $H(z)$ that correctly describes the cosmic acceleration is difficult to find as an $n$th order equation would have $n$ branches of $H(z)$.
	
	\subsection{$\Lambda$CDM}\label{model:1}
	As we want our models to have a $\Lambda$CDM limit, once a form of the $f(Q,T)$ function equivalent to $\Lambda$CDM is found, we propose extensions that {will have $\Lambda$CDM as a sub-case}. We can recover the $\Lambda$CDM density content by either introducing dark energy in a cosmological constant $\Lambda$ to the $f(Q,T)$ function or by establishing an effective equation of state.  An effective equation of state (EoS) of $w=-\dfrac{1}{1+m(1+z)^n}$ (with $w=p/\rho$)  drives the late-time accelerated expansion of the Universe and a dark matter-like evolution at early times. Starting with the {first} approach, let the $f(Q,T)$ function be
	\begin{equation}
		f(Q,T) = - \frac{1}{G_N} (Q + c Q^{1/2} + 2 \Lambda) ,
	\end{equation}
	with $G_N$ as the Newton's gravitational constant, $c$ as a constant parameter and $\Lambda$ as the cosmological constant. By taking this form of $f(Q,T)$, equations (\ref{eqn:standardFirstFriedmannEquation}) and (\ref{eqn:standardSecondFriedmannEquation}) become
	\begin{equation}
		\label{eqn:standardFriedmannEquations}
		\begin{split}
			H^2 = \frac{8 \pi G_N}{3} \rho + \frac{\Lambda}{3} \\
			\dot{H} + H^2 = - \frac{4 \pi G_N}{3} (\rho + 3p) + \frac{\Lambda}{3} ,
		\end{split}
	\end{equation}
	These are the $\Lambda$CDM Friedmann equations. As we can see, this form of $f(Q,T)$ gives the same background evolution of $\Lambda$CDM, however the $\Lambda$CDM perturbations are recovered only if $c=0$ \cite{jimenez2020cosmology}. Moreover, the energy balance equation is
	\begin{equation}
		\label{eqn:balanceEnergyEquation}
		\dot{\rho} + 3H(\rho + p) = 0 .
	\end{equation}
	
	In this way, we recover GR. To have the solution of $H(z)$ corresponding to $\Lambda$CDM, we need to consider the $\Lambda$CDM content of the Universe, i.e., baryonic matter plus cold dark matter plus radiation (dark energy is included in the cosmological constant). Baryonic matter and cold dark matter have both an EoS of $w=0$ and then from equation (\ref{eqn:balanceEnergyEquation}) we get $\rho_b = \rho_{0b} (1+z)^3$ and $\rho_{CDM} = \rho_{0 CDM} (1+z)^3$. Radiation has an equation of state $w=1/3$ and then $\rho_{r} = \rho_{0r} (1+z)^4$. The final content is dark energy whose density is $\rho_\Lambda = \dfrac{\Lambda}{8\pi G}$. Finally, considering the cosmological density parameters $\Omega_i = \rho_{0i}/\rho_{0cr}$ with $\rho_{0cr} = (3 H_0^2)/(8\pi G_N)$ the nowadays critical density, we get from the first Friedmann equation (\ref{eqn:standardFriedmannEquations})
	\begin{equation}
		\label{eqn:standarHubbleParameter}
		H(z) = H_0 \left[ (\Omega_b + \Omega_{CDM} ) (1+z)^3 + \Omega_r (1+z)^4 + \Omega_\Lambda \right]^{1/2} .
	\end{equation}
	
	Setting $z=0$ and then $H(0) = H_0$ we get the constrain
	\begin{equation}
		\label{eqn:cosmologyConstrain}
		\Omega_b + \Omega_{CDM} + \Omega_r + \Omega_\Lambda = 1 .
	\end{equation}
	
	The last equation implies that not all the parameters are independent. To simplify the equations we considered cold dark matter and baryonic matter together as matter ($\Omega_M = \Omega_b+\Omega_{CDM}$) because they both have the same background evolution. Furthermore, as we are working with low-redshifts, we took $\Omega_r=0$ because it is negligible in this redshift range. Therefore, the constrain that we consider is $\Omega_M + \Omega_\Lambda = 1$\\
	
	Another way to have $\Lambda$CDM as a sub-model is to introduce an effective EoS by assuming a perfect fluid with EoS $w=-\dfrac{1}{1+m(1+z)^3}$. In this case we do not need to introduce the cosmological constant in $f(Q,T)$, and instead we write $f(Q,T) = - (Q + c Q^{1/2})/G_N$. By taking this form of $f(Q,T)$ into equations (\ref{eqn:standardFirstFriedmannEquation}) and (\ref{eqn:standardSecondFriedmannEquation}) we recover equations (\ref{eqn:standardFriedmannEquations}) without the $\Lambda$ terms. However, the cosmological constant in this case is incorporated through the stress-energy tensor of the universe using the effective EoS that includes it and the dust matter. To match this with $\Lambda$CDM we need to do a variable change $\Omega_\Lambda = \dfrac{1}{1+m}$. In this way the $H(z)$ is the same as (\ref{eqn:standarHubbleParameter}) without the radiation term, negligible at low-redshifts. \\
	
	The models considered in this article are extensions of either $f(Q,T) = -(Q + c Q^{1/2} + 2\Lambda)/G_N$ or $f(Q,T) = -(Q + c Q^{1/2})/G_N$. We use $c=0$ in models \ref{model:2},\ref{model:3} and \ref{model:6} to reduce the number of parameters. Extensions of $f(Q,T) = -(Q + c Q^{1/2} + 2\Lambda)/G_N$ take an EoS ({{which is not an effective EoS}}) of $w=0$, i.e., they consider a pressure-less dust matter plus dark energy in the cosmological constant. Extensions of $f(Q,T) = -(Q + c Q^{1/2})/G_N$ have a negative effective EoS of $w = -\dfrac{1}{1+m(1+z)^n}$ which drives the late-times accelerated expansion of the Universe.\\
	
	In the extensions of $f(Q,T) = -(Q+c Q^{1/2}+2\Lambda)/G_N$, we considered a universe composed by dust matter, and therefore $w$ and $p$ are 0. This model replicates $\Lambda$CDM more accurately  as a sub-model because it considers both dark matter and a cosmological constant. \\
	
	On the other hand, with extensions of $f(Q,T) = -(Q + c Q^{1/2})/G_N$ we took the stated form of the EoS to reproduce the accelerated expansion at late-times. 
	
	\subsection{$f(Q,T) = -Q/G_N + b T$}\label{model:2}
	
	The first model that we considered here is given by
	\begin{equation}
		f(Q,T) = -\dfrac{Q}{G_N} + b T
	\end{equation}
	Where $b$ is a constant parameter. When $b=0$ this model has the field equations of $\Lambda$CDM (without the term $\Lambda$). By solving equations (\ref{eqn:firstFriedmannEquation}) and (\ref{eqn:secondFriedmannEquation}) for $p$,$\rho$ and taking their ratio, we get the equation of state 
	\begin{equation}
		\label{eqn:equationOfStateModel1}
		w = \frac{3H^2(8\pi + b) + \dot{H} (16 \pi + 3b)}{b \dot{H} - 3H^2 (8\pi + b)} .
	\end{equation}
	
	To solve this equation we need to take an ansatz on the form of $w$ and because of the discussion made in the last section, this EoS should be dark energy-like (at late-times). Following \cite{arora2021constraining} we considered a parametric form of the equation of state given by \cite{mukherjee2016acceleration}
	\begin{equation}
		\label{eqn:parametricEquationOfState}
		w = - \frac{1}{1+m(1+z)^n} .
	\end{equation}
	
	{This form of the EoS gives the overall behaviour of the Universe including both a dark energy component and matter. At low redshift, it reproduces the accelerated expansion of the Universe with $w < -1/3$ while at high redshifts it reproduces the early dark matter behaviour with $w \to 0$. As a result, this parametric form of $w$ constitutes an effective equation of state}. {{It is important to point out that this EoS is the mechamism that provides the accelerated expansion of the universe and not the $f(Q,T)$ function. Therefore, it also works for GR as \cite{mukherjee2016acceleration} pointed out}}. By plugging this form into equation (\ref{eqn:equationOfStateModel1}) and solving the differential equation, we get
	\begin{equation}
		H(z) = H_0 \left[ \frac{b+(16\pi+3b)(1+m(1+z)^n)}{b+(16\pi+3b)(1+m)} \right]^a ,
	\end{equation}
	Where $a = \dfrac{3(8\pi+b)}{n(16\pi+3b)}$. If we set $b=0$ and $n = 3$ the equation becomes
	\begin{equation}
		H(z) = H_0 \left[ \frac{1 + m(1+z)^3}{1+m} \right]^{1/2} ,
	\end{equation}
	Finally, by setting $\Omega_\Lambda = \dfrac{1}{1+m}$, we get
	\begin{equation}
		H(z) = H_0 \left[ (1-\Omega_\Lambda) (1+z)^3 + \Omega_\Lambda \right]^{1/2} .
	\end{equation}
	
	This function replicates the Hubble function of $\Lambda$CDM with $b=0$ and $n=3$ if we consider a Universe containing matter, a cosmological constant and a negligible radiation. Any deviation from this values of $b$ and $n$ will imply a deviation from $\Lambda$CDM. Applying a change of variable $\Omega_\Lambda = \dfrac{1}{1+m}$, the parameter vector for this model is $\Theta = \{ h, \Omega_\Lambda, b, n \}$.
	
	\subsection{$f(Q,T) = -(Q + 2 \Lambda)/G_N + bT$}\label{model:3}
	
	This function is the simplest generalization of $f(Q) = -(Q+2\Lambda)/G_N$ to an $f(Q,T)$ function and it reduces to $\Lambda$CDM when $b = 0$. Moreover, it has only one more parameter than $\Lambda$CDM. For simplicity, we will write provisionally $f(Q,T) = \Tilde{a} Q + bT + \Tilde{c}$ with $\Tilde{c} = -2\Lambda/G_N , \, \Tilde{a} = 1/G_N$. We considered that the Universe is made of dust pressure-less matter to use $w = 0$ and $p=0$. Setting $p=0$ and solving both equations (\ref{eqn:standardFirstFriedmannEquation}) and (\ref{eqn:standardSecondFriedmannEquation}) for the density 
	\begin{equation}
		\begin{split}
			\rho = \frac{\Tilde{c} - 6 \Tilde{a} H^2}{16\pi + 3b} \\
			\rho = \frac{8 \Tilde{a} \dot{H} + 6 \Tilde{a} H^2 - \Tilde{c}}{16\pi + b} .
		\end{split}
	\end{equation}
	
	By equating both and by using $\dot{H} = -(1+z) H dH/dz$ we get
	\begin{equation}
		\frac{dH}{dz} = \frac{8\pi + b}{16\pi +3b} \frac{6 \Tilde{a} H^2 - \Tilde{c}}{2\Tilde{a} H(1+z)}.
	\end{equation}
	
	Now, we solve the equation and return to the original variables to arrive to
	\begin{equation}
		\label{eqn:Hzmodel3}
		H(z) = \left[\frac{\Lambda}{3} + \left(H_0^2 - \frac{\Lambda}{3} \right) (1+z)^\lambda \right]^{1/2} .
	\end{equation}
	
	This form can be further simplified if we use the change of variable $\Lambda = 3H_0^2 \Omega_\Lambda$
	\begin{equation}
		H(z) = H_0 \left[ \Omega_\Lambda + (1-\Omega_\Lambda ) (1+z)^\lambda \right]^{1/2} ,
	\end{equation}
	Where $\lambda = 6(8\pi+b)/(16\pi+3b)$. By setting $b = 0$ and considering that $\Omega_M + \Omega_\Lambda = 1$ we recover the $\Lambda$CDM Friedmann equations. As we can see, this model modifies the exponent in the matter component. The parameter vector for this model is $\Theta = \{ h, \Omega_\Lambda, b \}$
	
	\subsection{$f(Q,T) = -(Q+2H_0^2 c (Q/(6H_0^2))^{n+1})/G_N + bT$}\label{model:4}
	
	This model comes from  $f(Q,T) = -Q/G_N + bT$ and it can be seen as an extension of the $f(Q)$ model given by \cite{khyllep2021cosmological} where it was studied at a background and at a perturbative context. It was also studied by \cite{ayuso2021observational}. Now, we are considering an $f(Q)$ function given by
	
	\begin{equation}
		f(Q) = \frac{1}{8 \pi G_N} \left[ Q - 6 \lambda M^2 \left( \frac{Q}{6M^2} \right)^n \right] .
	\end{equation} 
	
	This means that this $f(Q,T)$ model gives $\Lambda$CDM for $b=0$ and $c = 0$, or $n = -1/2$. {However we will focus on the integer values of $n$ because $b$ and $c$ will be the parameters accounting deviations from the $\Lambda$CDM model}. We solve the equations (\ref{eqn:standardFirstFriedmannEquation}) and  (\ref{eqn:standardSecondFriedmannEquation}) for $p$ and $\rho$ to take their ratio and to get the EoS
	\begin{equation}
		w = \frac{6H^2 (8\pi + b) + 2c (2n+1) H^{2n+2}/H_0^{2n} (8\pi + b) + 2b \dot{H} [1 + 1/3 c (n+1) (2n+1) H^{2n} /H_0^{2n}] (16\pi + 3b)}{2b\dot{H} [1+2 c (n+1)(2n+1) H^{2n} /H_{0}^{2n}] - 6H^2(8\pi+b) - 1/3 c (2n+1)H^{2n+2}/H_0^{2n} (8\pi + b) } .
	\end{equation}
	
	Similarly as applied to the model $f(Q,T) = -Q/G_N + bT$, we considered a parametric equation of state given by \cite{mukherjee2016acceleration}
	\begin{equation}
		w = - \frac{1}{1+m(1+z)^p} 
	\end{equation}
	Where we wrote the exponent of $1+z$ as $p$ to avoid confusing notation. {Equating} both equations of state and solving for $dH/dz$ we get the differential equation
	\begin{equation}
		\frac{dH}{dz} = \frac{8\pi+b}{2} \frac{[6H^2 + 2c (2n+1) H^{2n+2}/H_0^{2n}]m(1+z)^{p-1}}{[1+1/3 c (n+1)(2n+1) H^{2n+1} /H_0^{2n}] [b+(16\pi+3b)(1+m(1+z)^p)]} ,
	\end{equation}
	By solving it in terms of the dimensionless Hubble parameter $E(z) = H(z)/H_0$, {we get}
	\begin{equation}
		\label{eqn:sol34}
		\frac{(2n+1)c}{3}E^{2n+2}(z) + E^2(z) = \left[1+\frac{(2n+1)c}{3} \right] \left[ \frac{b+(16\pi+3b)(1+m(1+z)^p)}{b+(16\pi+3b)(1+m)} \right]^\lambda ,
	\end{equation}
	Where $\lambda = \dfrac{6(8\pi+b)}{p(16\pi+3b)}$. Setting $b=0$, $p=3$ and $c=0$ or $n=-1/2$ we get
	\begin{equation}
		H(z) = H_0 \left[ \frac{1+m(1+z)^3}{1+m} \right]^{1/2} .
	\end{equation}
	
	We mimic the Hubble function of $\Lambda$CDM if we define $\Omega_\Lambda = \dfrac{1}{1+m}$. {To solve the equation, we need to choose a value for $n$. By setting $n=0$ we would recover the solution of the model 3.2. If we set $n=1$ or $n=-2$, we get a quadratic equation of $E^2(z)$. However, with $n=1$ the solution is ({after we perform} a variable change of $p \to n$ to be consistent with $f(Q,T) = -Q/G_N + bT$)}
	\begin{equation}
		E^2(z) = \frac{-1 + \left[ 1 + 4c(1+c) \left( \dfrac{b+(16 \pi +3b)(1+m(1+z)^n)}{b+(16 \pi + 3b)(1+m)} \right)^\lambda \right]^{1/2}}{2c}. 
	\end{equation}
	
	This is the positive branch of the quadratic equation, the negative one would predict a decelerating universe, in contradiction with current observations. This solution, however, {did not include the value of the singularity $c=0$ inside the confidence levels} and subsequently did not contain $\Lambda$CDM as a sub-model. Therefore, following \cite{ayuso2021observational,jimenez2020cosmology} , we will solve equation (\ref{eqn:sol34}) by taking $n=-2$ and after this substitution, we make the variable change $p\to n$ to be consistent with model $f(Q,T) = -Q/G_N + bT$ 
	\begin{equation}
		E^2(z) = \frac{(1-c) \left[ \dfrac{b+(16\pi+3b)(1+m(1+z)^n)}{b+(16\pi+3b)(1+m)} \right]^\lambda + \left[ (1-c)^2 \left[ \dfrac{b+(16\pi+3b)(1+m(1+z)^n)}{b+(16\pi+3b)(1+m)} \right]^{2\lambda} + 4c \right]^{1/2}}{2} .
	\end{equation}
	
	Where $\lambda = \dfrac{6(8\pi+b)}{n(16\pi+3b)}$. With $\Omega_\Lambda = \dfrac{1}{1+m}$, the parameter vector of this model is $\Theta = \{ h, \Omega_\Lambda, b, c, n \}$. 
	
	In the case that $n=-2$, {it is easy to choose the right branch because} it has a quadratic equation. The negative branch would give $E<0$ if $c>0$ which means a decelerating Universe, in contradiction with current observations. Models with a different $n$ {would give a higher order equation. For example a third order equation would have three different branches of $H(z)$, and in that case it would be hard to know the branch of $H(z)$ that is consistent with the data. This is the reason why we solved the equation with $n=-2$}.
	
	\subsection{$f(Q,T) = -(Q+2 H_0^2 c (Q/(6H_0^2))^{n+1} + 2\Lambda)/G_N + bT$}\label{model:5}
	
	This model can also be seen as an extension of the $f(Q)$ model studied by \cite{khyllep2021cosmological} and  \cite{ayuso2021observational}. The $b$ parameter drives the transition from $f(Q)$ to $f(Q,T)$ and by setting $c=0$ and $b=0$ we recover GR. By taking $p=0$, i.e., $w=0$ and solving equations (\ref{eqn:standardFirstFriedmannEquation}) and (\ref{eqn:standardSecondFriedmannEquation}), we get the differential equation
	\begin{equation}
		\frac{dH}{dz} = \frac{8\pi + b}{16\pi + 3b} \frac{6H^2 + 2 c (2n+1) H^{2n+2} /H_0^{2n} - 2\Lambda}{2(1+z)(1+c(n+1)(2n+1) H^{2n}/H_0^{2n}/3)H} .
	\end{equation}
	
	To be consistent with the parameters of $f(Q,T) = -(Q+2\Lambda)/G_N + bT$ we redefine $\Lambda = 3H_0^2 \Omega_\Lambda$ and rewrite the equation in terms of the dimensionless Hubble parameter $E(z) = H(z)/H_0$, we get
	\begin{equation}
		\label{eqn:generalEc5}
		\frac{(2n+1)c}{3} E^{2n+2}(z) + E^2(z) = \Omega_\Lambda + \left[(1-\Omega_\Lambda) + \frac{c(2n+1)}{3} \right] (1+z)^\lambda ,
	\end{equation}
	With $\lambda = 6(8 \pi + b)/(16 \pi + 3b)$. As we can see, by setting $c=0$ and  $ b=0$ or $n=-1/2$ we recover $\Lambda$CDM. To solve equation (\ref{eqn:generalEc5}) a value of $n$ needs to be taken. As well as with the previous model (because this one also {{does not include the value of the singularity $c= 0$ inside the confidence levels, and therefore it does not include $\Lambda$CDM}}) we will take $n=-2$, and hence
	\begin{equation}
		E^2(z) = \frac{\Omega_\Lambda+ \left[(1-\Omega_\Lambda)-c\right](1+z)^\lambda + \left[ \left(\Omega_\Lambda+ [(1-\Omega_\Lambda)-c](1+z)^\lambda\right)^2 + 4c \right]^{1/2}}{2} .
	\end{equation}
	
	Here the positive branch was taken because the negative one with $c>0$ predicts a decelerating universe, in contradiction with the observations. For this model, the vector of parameters is $\Theta = \{ h, \Omega_\Lambda, b, c \}$.\\
	
	Other values of $n$ can be taken to test this model against $\Lambda$CDM. However, $n=-2$ and $n=1$ are the simplest values because they both have a quadratic equation of the dimensionless Hubble parameter $E(z)$. Models with a different value of $n$ would have an equation of higher order. This would complicate the selection of the right branch {of $H(z)$, just as in the previous model}. 
	
	\subsection{$f(Q,T) = -(Q+2\Lambda)/G_N - ((16\pi)^2 G_N b)/(120 H_0^2) T^2$} \label{model:6}
	
	This model modifies the $f(Q,T) = aQ -bT^2$ model of \cite{xu2019f, godani2019frw} adding a cosmological constant to reproduce the accelerated expansion of the universe. By solving equation (\ref{eqn:standardFirstFriedmannEquation}) in terms of the density (taking the positive branch of the quadratic solution since the density is positive) and introducing the result into equation (\ref{eqn:evolutionH}), we get the differential equation of the dimensionless Hubble parameter
	\begin{equation}
		\frac{dE}{dz} = \frac{3}{b E(z) (1+z)} \left[ \left( 1 + b(E^2(z)-\Omega_\Lambda)\right)^{1/2} - 1\right]  \left[ 1 + \dfrac{2 \left(\left( 1 + b(E^2(z) - \Omega_\Lambda) \right)^{1/2} - 1\right)}{5} \right] 
	\end{equation}
	
	This equation can be handled numerically. For this model our vector of parameters is given by $\Theta = \{ h, \Omega_\Lambda, b \}$.

\section{Observational Datasets} \label{sec:obsdatabase}

To constrain the vector of parameters $\Theta$ of the $f(Q,T)$ models presented in the previous section, we use two observational datasets: the Cosmic Chronometers (CC) compilation \cite{magana2018cardassian} which gives values of the Hubble parameter $H(z)$, and the Pantheon type Ia Supernovae compilation (SNeIa) \cite{scolnic2018complete} that gives values of the distance modulus $\mu(z)$. 

\subsection{Cosmic Chronometers + $H_0$ (CC+$H_0$)}

This dataset is the compilation presented in \cite{magana2018cardassian}. Cosmic Chronometers are $N_{\text{CC}} = 51$ measurements of the Hubble parameter $H(z)$ along with their uncertainties in the redshift range $z \in (0, 2.36]$. Aside from that, we consider the SH0ES collaboration's \cite{Riess_2019} Hubble constant. We named the two datasets together as CC+$H_0$. 
{To measure the Hubble parameter, we consider an FLRW universe through $H(z) = -\dfrac{1}{1+z} \dfrac{dz}{dt}$. We get the term $\dfrac{dz}{dt}$ using the differential age evolution of early-type galaxies \cite{jimenez2002constraining, Nunes_2017}. This shows that we can compute the Hubble parameter in a model-independent way because, as it is known, the SH0ES collaboration's value of $H_0$ is also model-independent.} To get the best fit parameters of our models, we minimize the $\chi^2$ function defined as
\begin{equation}
	\label{eqn:chiSquaredCC}
	\chi_{\text{CC}}^2 = \sum_{n=1}^{N_{\text{CC}}} \left( \dfrac{H(\Theta, z_n)-H_{n \, \text{obs}}}{\sigma_{n \, \text{obs}}} \right)^2 ,
\end{equation}
where $\Theta$ is the vector containing the parameters of the model, $H(\Theta, z_n)$ the theoretical Hubble parameter that depends on the model, $H_{n \, \text{obs}}$ the Hubble parameter reported in the CC+$H_0$ compilation, and $\sigma_{n \, \text{obs}}$ the uncertainty of the Hubble parameter. 

\subsection{Pantheon SNeIa + $H_0$}

The next dataset considered is the Pantheon compilation \cite{scolnic2018complete}, consisting of $N_{\text{Pantheon}} = 1048$ measurements of the distance modulus along with their uncertainties in the redshift range $z \in [0.010, 2.26]$. Similarly to the CC+$H_0$ case we got the best fit parameters by minimizing the $\chi^2$ function, in this case it is defined in terms of the distance modulus $\mu(z)$ instead of $H(z)$
\begin{equation}
	\label{eqn:chiSquaredPantheon}
	\chi_{\text{Pantheon}}^2 = \sum_{n=1}^{N_{\text{Pantheon}}} \left( \dfrac{\mu(\Theta, z_n)-\mu_{n \, \text{obs}}}{\sigma_{n \, \text{obs}}} \right)^2 ,
\end{equation}
where $\Theta$ is the vector containing the parameters of a model, $\mu(\Theta,z_n)$ the theoretical distance modulus (it depends on the model considered), and $\mu_{n \, \text{obs}}$ the distance modulus {of the Pantheon compilation calibrated with the $H_0$ value from SH0ES \cite{Riess_2019}} . {Due to the calibration of the data with this value of $H_0$ we will call this dataset Pantheon+$H_0$}. Since we presented the theoretical Hubble parameter of our models in section \ref{sec:models}, we need to write the distance modulus in terms of this Hubble parameter. We start by writing the theoretical distance modulus in terms of the luminosity distance
\begin{equation}
	\label{eqn:distanceModulus-HubbleParameter}
	\mu(\Theta, z) = 5 \, \log_{10} \left( \frac{D_L(\Theta, z)}{10 \, \text{pc}} \right) ,
\end{equation}
where $D_L$ is the theoretical luminosity distance given by
\begin{equation}
	\label{eqn:luminosityDistance}
	D_L(\Theta, z) = \frac{c(1+z)}{H_0} \int_{0}^z \frac{dz'}{E(\Theta, z')} .
\end{equation}

$E(\Theta, z)$ is the dimensionless Hubble parameter  $E(z)=H(z)/H_0$. Therefore, we can test our models with the Pantheon compilation too. We will call the joint CC+$H_0$ and Pantheon\textbf{+$H_0$} compilations Pantheon+CC+$H_0$.

\section{Methodology}
\label{sec:methodology}

We used the \texttt{emcee} python package \cite{foreman2013emcee} to perform the Markov Chain Monte Carlo (MCMC) method. After minimizing the $\chi^2$ function, we initialize 32 random walkers around a Gaussian ball of the best fit vector for the parameters $\Theta_{\text{best fit}}$, as suggested by \cite{foreman2013emcee}. Once the MCMC was done, we used the \texttt{GetDist} python package  \cite{lewis2019getdist} to compute the regions with their 68.3\% and 95\%  confidence levels and the best fit MCMC values with their  68.3\% confidence levels. \\

Regarding the confidence regions of the parameters $\Theta$ of each model, we present one triangle plot per model containing an analysis performed with the following steps
\begin{enumerate}
	\item Analysis with the CC+$H_0$ compilation
	\item Analysis with the Pantheon+$H_0$ compilation
	\item Analysis with the joint Pantheon+CC+$H_0$ compilation
\end{enumerate}

The best fits of the parameters $\Theta$ of each model along with their uncertainties are in table \ref{tab:resultsBestFits}, and their confidence regions in Fig. \ref{fig:trianglePlots}. \\

To test our models against the $\Lambda$CDM model, we used the Bayes evidence criterion. This criterion is used to compute the preference of a model over another. To use this criterion we need to compute the evidence of a model 
\begin{equation}
	\mathcal{E} = \int d\Theta \, \theta(\Theta|\mathcal{M}) \, \mathcal{L}(D|\Theta, \mathcal{M}) ,
\end{equation}
where $\mathcal{M}$ is the model, $\theta(\Theta|\mathcal{M})$ is the prior probability of the model parameters, and $\mathcal{L}(D|\Theta, \mathcal{M})$ is the Likelihood \cite{hogg2010data}. 

To compare between two models $\mathcal{M}_1$ and $\mathcal{M}_2$ we need to take the natural logarithm of the ratio of their evidences
\begin{equation}
	\label{eqn:bayesianRatio}
	\ln B_{12} = \ln \left(\frac{\mathcal{E}_1}{\mathcal{E}_2}\right) .
\end{equation}

Provided that $\mathcal{E}_1>\mathcal{E}_2$, $\mathbf{\mathcal{M}_1}$ is preferred and the strength of the preference is given by the Jeffrey’s criterion \cite{kass1995bayes,escamilla2016status} in table \ref{tab:Jeffrey'sBayesCriterion} . We call $\mathcal{M}_1$ the concordance model, and $\mathcal{M}_2$ the alternative models.

\begin{table}[ph]
\centering
{\begin{tabular}{@{}cc@{}} \toprule
Condition & Preference for $\mathcal{M}_1$ \\ \toprule
$0<\ln B_{12}<1$ & Weak \\ \colrule
$1<\ln B_{12}<2.5$ & Substantial \\ \colrule
$2.5<\ln B_{12}<5$ & Strong \\ \colrule
$\ln B_{12} >5$ & Very strong \\ \botrule
\end{tabular} \label{tab:Jeffrey'sBayesCriterion}}
\caption{Jeffrey's Bayes criterion to compare two models $\mathcal{M}_1$ and $\mathcal{M}_2$. If $\mathcal{E}_1>\mathcal{E}_2$ the $\mathcal{M}_1$ is preferred and the strength of the preference is given by this criterion}
\end{table}

\section{Analysis} \label{sec:model}
We computed the natural logarithm of the evidence of our models with the joint Pantheon+CC+$H_0$ compilation and the \texttt{dynesty} nested sampler python package \cite{speagle2020dynesty, skilling2004nested, skilling2006nested, feroz2009multinest} . These results are reported in table \ref{tab:criteriaComparison}, along with the comparison of the five models against $\Lambda$CDM. \\

As we can see in the triangle plots \ref{fig:trianglePlots}  and in the best fit values of the parameters with their confidence levels {(C.L.)} \ref{tab:resultsBestFits}, our alternative {{models}} mimic {the $\Lambda$CDM behaviour (when $b=0$ and $c=0$)} within the 95\% C.L. Thus, they have a $\Lambda$CDM limit. \\

We can also compare the Hubble constant computed with Pantheon+CC+$H_0$ with the one of the SH0ES \cite{Riess_2019} collaboration. To compute the tension between different values, we use \cite{camarena2018impact,cruz2021late}
\begin{equation}
	\label{eqn:tension}
	T_{H_0} = \frac{|H_{0 \, \text{model 1}} - H_{0 \, \text{model 2}}|}{(\sigma_{H_0 \, \text{model1}}^2 + \sigma_{H_0 \, \text{model2}}^2)^{1/2}}.
\end{equation}
Now, we focus our attention on table \ref{tab:resultsBestFits}. Using equation (\ref{eqn:tension}) to calculate the tension between our values and the SH0ES value of the Hubble constant $H_0 = 73.5 \, \pm \, 1.4\ \, \text{km} \, \text{s}^{-1} \, \text{Mpc}^{-1} $. This calculation yields to the result that our models lie {within the 95\% C.L. of the SH0ES value \cite{Riess_2019}. {Thus}, the Hubble constant's values of {all} our models {are in agreement} with the value of the SH0ES collaboration, {as expected}.\\
	
	{As a possible future work, we can test whether these models alleviate the tension between the late and early time data using the Planck analysis \cite{aghanim2020planck}. In particular, following the procedure of Dainotti et. al. \cite{dainotti2021hubble} to calculate the tension between Planck and late-time data for the $\Lambda$CDM model and $w_0 w_a$CDM models using Pantheon data. }\\
	
	Now, according to table \ref{tab:criteriaComparison}, $\Lambda$CDM is the best model compared to all the alternative models except to the last one $f(Q,T) = -(Q+2\Lambda)/G_N - ((16\pi)^2 G_N b)/(120 H_0^2) T^2$ because it has the highest value of the evidence. As stated by the Jeffrey's criterion (see table \ref{tab:Jeffrey'sBayesCriterion}), the preference for the concordance model is weak against $f(Q,T) = -Q/G_N + bT$ and $-(Q+2\Lambda)/G_N + bT$, substantial against$f(Q,T) = -(Q+2H_0c(Q/(6H_0^2))^{n+1})/G_N + bT$, and it is strong against $f(Q,T) = -(Q+2H_0c(Q/(6H_0^2))^{n+1} + 2\Lambda)/G_N + bT$ . However, $f(Q,T) = -(Q+2\Lambda)/G_N - ((16\pi)^2 G_N b)/(120 H_0^2) T^2$ shows a {substantial} preference against the $\Lambda$CDM concordance model. Hence, it deserves deeper future studies with other datasets such as the Cosmic Microwave Background (CMB) or Large-Scale Structure (LSS) to test if this model is a serious alternative to the standard cosmological model. \\
	
	Apart from the previous ideas, a perturbative approach is crucial to see the cosmological implications of $f(Q,T) = -(Q+2\Lambda)/G_N - ((16\pi)^2 G_N b)/(120 H_0^2) T^2$ and to analyze if the model can challenge $\Lambda$CDM at the perturbative level, aside from the background perspective presented here.

\begin{table}[H]
\centering
{\begin{tabular}{@{}ccccccc@{}} \toprule
Model & Compilation & $h$ & $\Omega_\Lambda$ & $b$ & $c$ & $n$  \\ \toprule
\multirow{3}{*}{1} & CC+$H_0$ & $0.7149\pm 0.0098$ & $0.752^{+0.015}_{-0.014}$ & - & - & - \\
& Pantheon+$H_0$ & $0.7284\pm 0.0023$ & $0.714\pm 0.013$ & - & - & - \\
& Pantheon+CC+$H_0$ & $0.7331\pm 0.0019$ & $0.7555\pm 0.0072$ & - & - & - \\ \colrule
\multirow{3}{*}{2} & CC+$H_0$ & $0.715\pm 0.015$ & $0.747^{+0.054}_{-0.044}$ & $1.83^{+3.1}_{-0.96}$ & - & $3.26^{+0.56}_{-0.64}$ \\
& Pantheon+$H_0$ & $0.7320\pm 0.0032$ & $0.771^{+0.044}_{-0.036}$ & $0.4^{+4.3}_{-2.1}$ & - & $4.39^{+0.97}_{-0.81}$ \\
& Pantheon+CC+$H_0$ & $0.7316\pm 0.0025$ & $0.731\pm 0.023$ & $1.8^{+3.2}_{-1.1}$ & - & $2.92\pm 0.35$ \\ \colrule
\multirow{3}{*}{3} & CC+$H_0$ & $0.707\pm 0.014$ & $0.714^{+0.054}_{-0.044}$ & $2.2^{+2.1}_{-3.2}$ & - & -\\ 
& Pantheon+$H_0$ & $0.7302\pm 0.0030$ & $0.763^{+0.064}_{-0.044}$ & $-3.4^{+1.8}_{-5.2}$ & - & - \\ 
& Pantheon+CC+$H_0$ & $0.7310\pm 0.0023$ & $0.723^{+0.023}_{-0.021}$ & $2.8^{+1.5}_{-2.2}$ & - & - \\ \colrule
\multirow{3}{*}{4} & CC+$H_0$ & $0.718\pm 0.014$ & $0.47\pm 0.26$ & $5.97^{+4.0}_{-0.66}$ & $0.284^{+0.34}_{-0.084}$ & $3.4^{+2.0}_{-1.1}$ \\
& Pantheon+$H_0$ & $0.7317\pm 0.0029$ & $0.46\pm 0.26$ & $0.9\pm 5.0$ & $0.310^{+0.33}_{-0.096}$ & $3.81^{+2.1}_{-0.65}$ \\
& Pantheon+CC+$H_0$ & $0.7325\pm 0.0025$ & $0.45^{+0.20}_{-0.34}$ & $5.4^{+4.6}_{-1.5}$ & $0.259^{+0.32}_{-0.082}$ & $2.4^{+1.5}_{-1.2}$ \\ \colrule
\multirow{3}{*}{5} & CC+$H_0$ & $0.714\pm 0.013$ & $0.38^{+0.13}_{-0.33}$ & $5.3^{+4.3}_{-1.7}$ & $0.28^{+0.26}_{-0.15}$ & - \\
& Pantheon+$H_0$ & $0.7309\pm 0.0028$ & $0.38^{+0.13}_{-0.38}$ & $-0.7^{+3.3}_{-6.3}$ & $0.30^{+0.28}_{-0.14}$ & - \\
& Pantheon+CC+$H_0$ & $0.7322\pm 0.0024$ & $0.44^{+0.17}_{-0.25}$ & $6.1^{+3.9}_{-1.6}$ & $0.23^{+0.20}_{-0.15}$ & - \\ \colrule
\multirow{3}{*}{6} & CC+$H_0$ & $0.699\pm 0.013$ & $0.698^{+0.047}_{-0.028}$ & $1.3839^{+0.0042}_{-1.4}$ & - & - \\
& Pantheon+$H_0$ & $0.7196\pm 0.0025$ & $0.673\pm 0.026$ & $6.7\pm 5.9$ & - & - \\
& Pantheon+CC+$H_0$ & $0.7230\pm 0.0022$ & $0.724^{+0.016}_{-0.012}$ & $1.04^{+0.13}_{-0.91}$ & - & - \\
\botrule
\end{tabular} \label{tab:resultsBestFits}}
\caption{Best fits of the parameters of our models with their 68.3\% confidence level using CC+$H_0$, Pantheon+$H_0$ and Pantheon+CC+$H_0$ compilations. Model 1 corresponds to $\Lambda$CDM, 2 to $f(Q,T) = -Q/G_N + bT$, 3 to $f(Q,T) = -(Q+2\Lambda)/G_N + bT$, 4 to $f(Q,T) = -(Q+2H_0^2 c (Q/(6H_0^2))^{n+1})/G_N + bT$, 5 to $f(Q,T) = -(Q+2 H_0^2 c (Q/(6H_0^2))^{n+1} + 2\Lambda)/G_N + bT$, and 6 to $f(Q,T) = -(Q+2\Lambda)/G_N - ((16\pi)^2 G_N b)/(120 H_0^2) T^2$.}
\end{table}

\begin{figure}[H]
    \centering
    \includegraphics[scale=0.60]{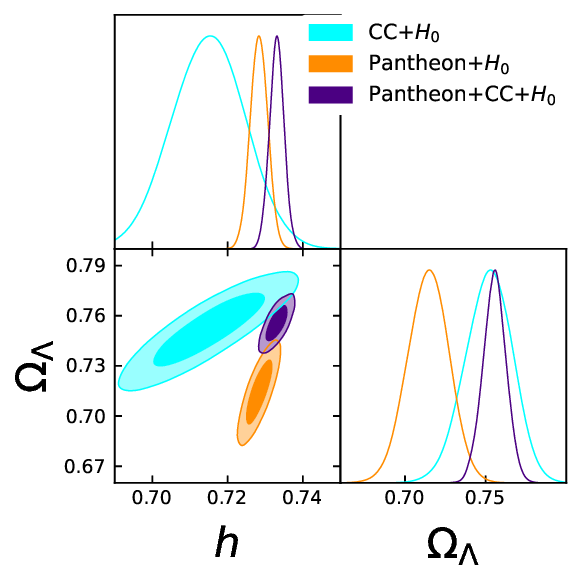}
    \includegraphics[scale=0.30]{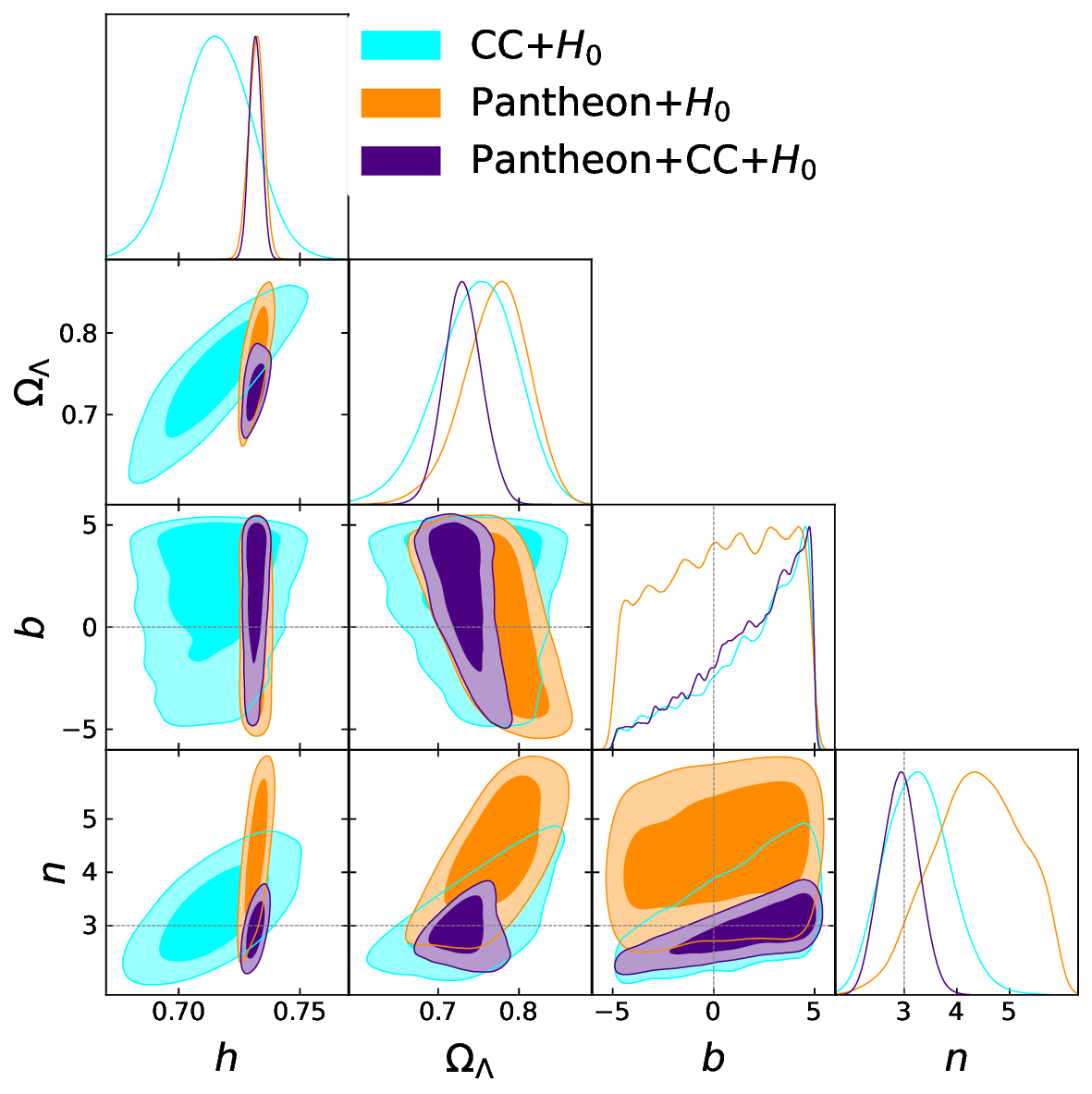}
    \includegraphics[scale=0.40]{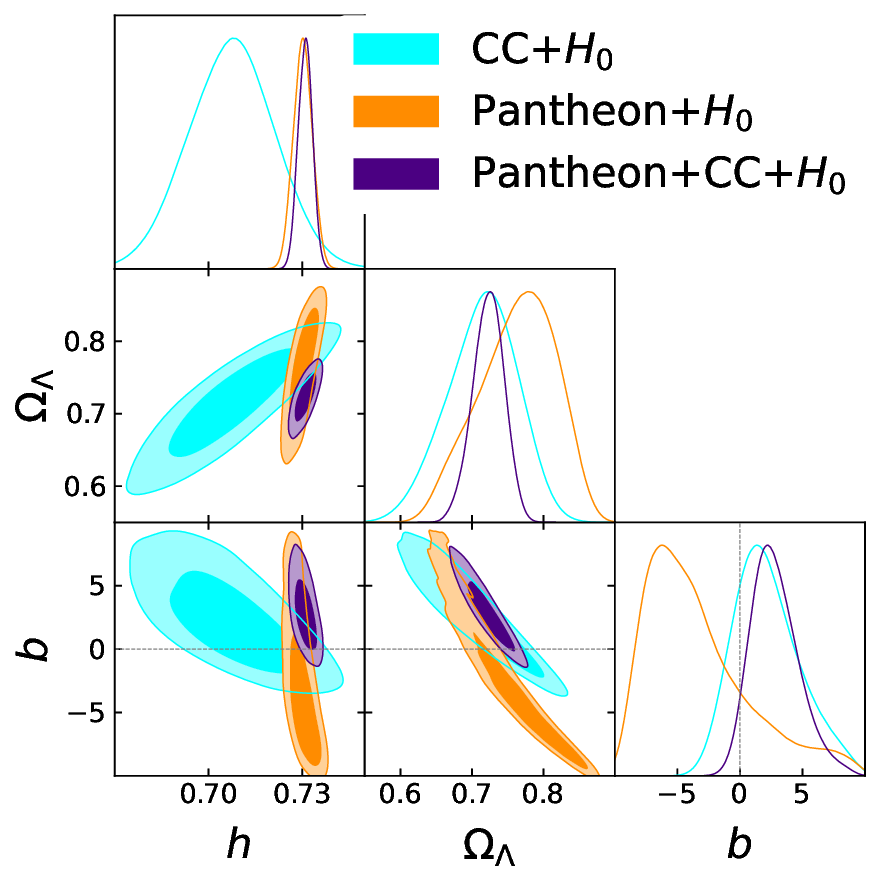}
    \includegraphics[scale=0.24]{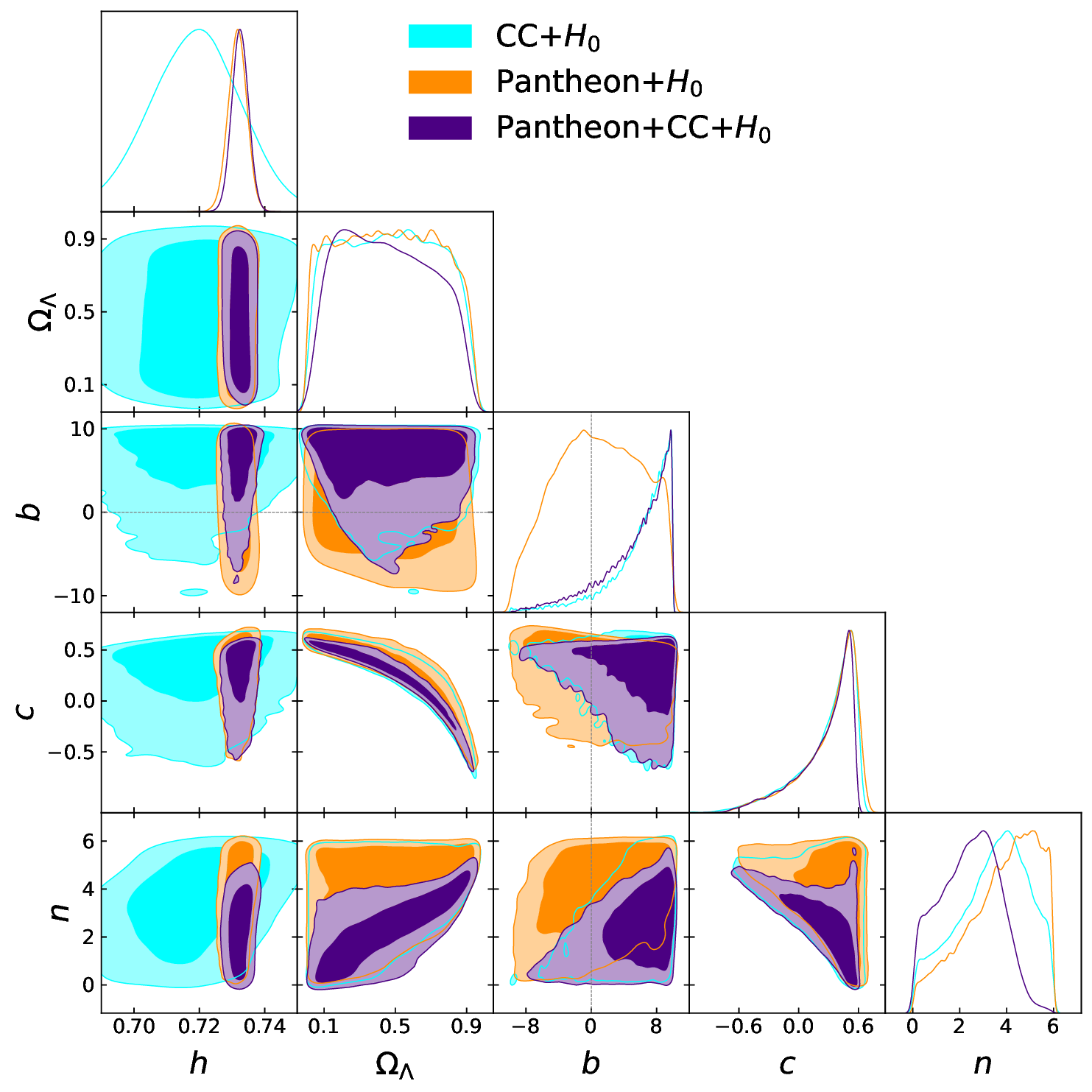}
    \includegraphics[scale=0.30]{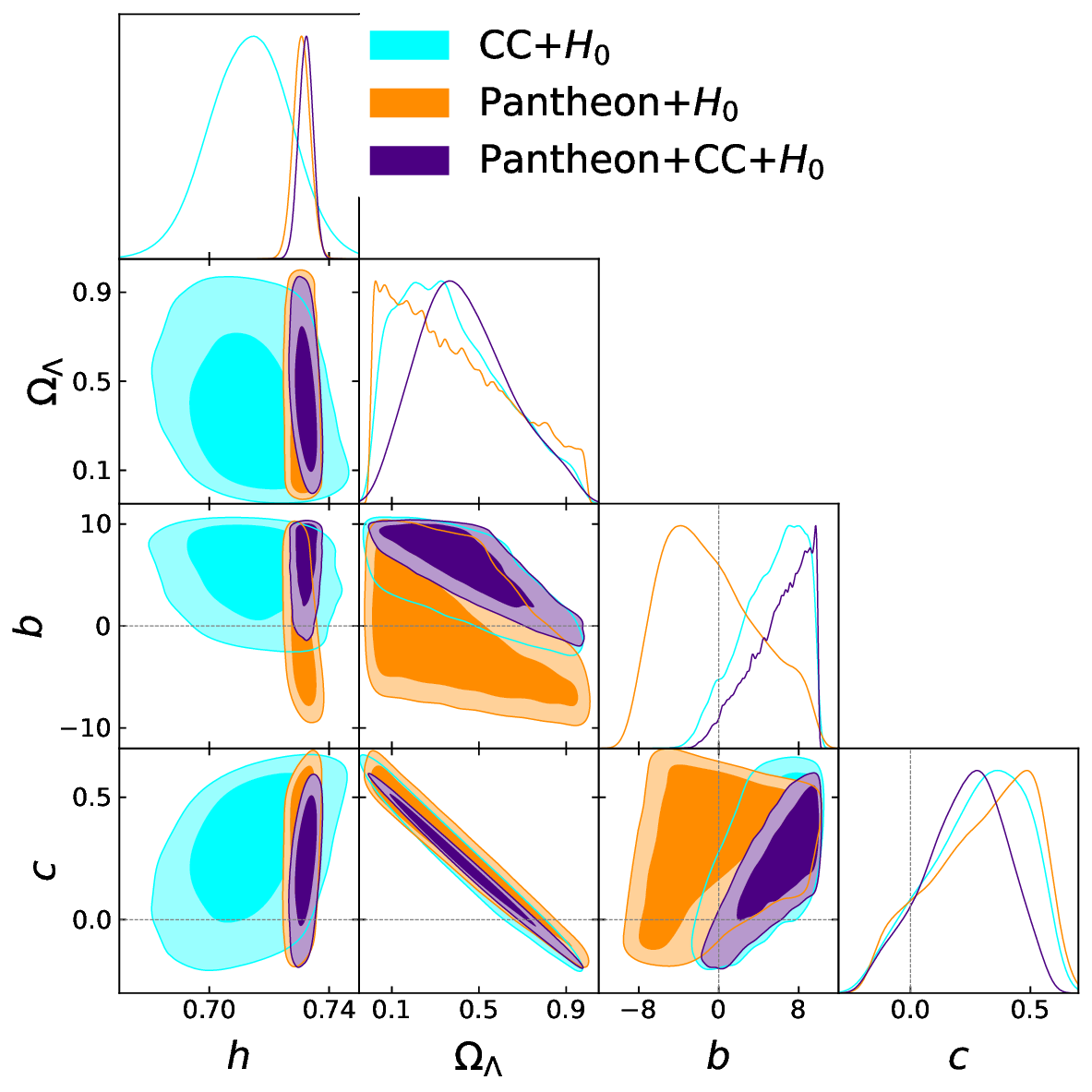}
    \includegraphics[scale=0.40]{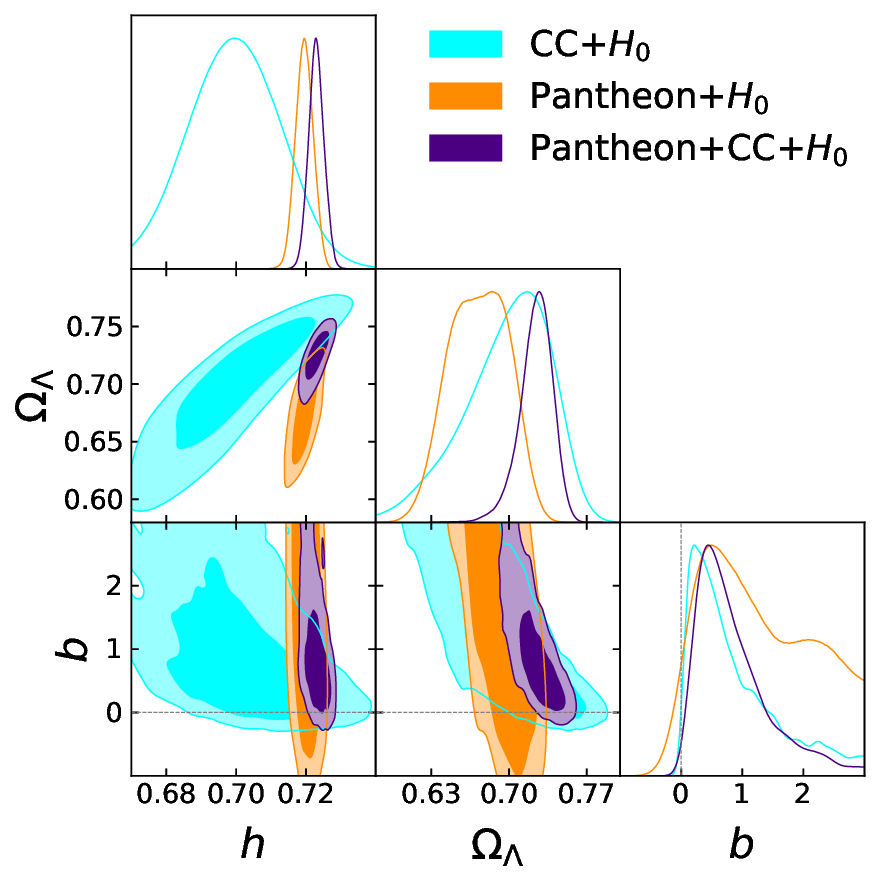}
    \caption{68.3\% and 95\% confidence levels of the parameters $\Theta = \{ h, \Omega_\Lambda, $b$, $c$, $n$ \}$ with the CC+$H_0$, Pantheon+$H_0$ and Pantheon+CC+$H_0$ compilations. Top left: $\Lambda$CDM, top right: $f(Q,T) = -Q/G_N + bT$, middle left: $f(Q,T) = -(Q+2\Lambda)/G_N + bT$, middle right: $f(Q,T) = -(Q+2H_0^2 c (Q/(6H_0^2))^{n+1})/G_N + bT$, bottom left:$f(Q,T) = -(Q+2 H_0^2 c (Q/(6H_0^2))^{n+1} + 2\Lambda)/G_N + bT$, bottom right: $f(Q,T) = -(Q+2\Lambda)/G_N - ((16\pi)^2 G_N b)/(120 H_0^2) T^2$.}
    \label{fig:trianglePlots}
\end{figure}

\begin{table}[H]
\centering
{\begin{tabular}{@{}ccc@{}} \toprule
$f(Q,T)$ & $\ln \mathcal{E}$ & $\ln B_{\Lambda CDM} \, \text{model} $ \\ \toprule
$\Lambda CDM$ & 386.22 & 0 \\ \colrule
$-Q/G_N + bT$ & 385.69 & 0.52 \\ \colrule
$-(Q+2\Lambda)/G_N+bT$ & 385.90 & 0.32 \\ \colrule
$-(Q+2 H_0 c (Q/(6H_0^2))^{n+1})/G_N + bT$ & 384.37 & 1.85 \\ \colrule
$-(Q+2H_0c(Q/(6H_0^2))^{n+1} + 2\Lambda)/G_N + bT$ & 382.89 & 3.33 \\ \colrule
$-(Q+2\Lambda)/G_N - ((16\pi)^2 G_N b)/(120 H_0^2) T^2$ & 387.80 & -1.58 \\ \botrule
\end{tabular} \label{tab:criteriaComparison}}
\caption{Comparison of the models presented in section \ref{sec:models} against $\Lambda$CDM using the Bayes Evidence criterion. A positive value indicates that $\Lambda$CDM is the preferred model whereas a negative value means that the alternative model is preferred.}
\end{table}

\section{Conclusions} \label{sec:Conclusions}

In the present article, we made use of symmetric teleparallel gravity mathematical description of gravity to propose and test five $f(Q,T)$ alternative models to the standard $\Lambda$CDM model using Cosmic Chronometers+$H_0$ and Pantheon Supernovae Ia + $H_0$ data. Our analyses are at a background perspective. {The cosmological parameters of our models lie within the 95\% C.L. of { {{the}} $\Lambda$CDM {{behaviour ($b=0$ and $c=0$)}}}. Moreover, the MCMC best fits of the Hubble constant lied within the 95\% C.L. of the SH0ES collaboration's value of $H_0 = 73.5 \, \pm \, 1.4 \, \text{km} \, \text{s}^{-1} \, \text{Mpc}^{-1} $ for all the models.\\
	
	The Bayesian evidence criterion showed us that $f(Q,T) = -(Q+2\Lambda)/G_N - ((16\pi)^2 G_N b)/(120 H_0^2) T^2$  {has a {substantial} preference against $\Lambda$CDM}. In regards to the remaining $f(Q,T)$ models, $\Lambda$CDM is the preferred one showing a weak preference against $f(Q,T) = -Q/G_N + bT$ and $f(Q,T) = -(Q + 2\Lambda)/G_N + bT$, a substantial one against $f(Q,T) = -(Q+2H_0c(Q/(6H_0^2))^{n+1})/G_N + bT$ and a {strong preference against} $f(Q,T) = -(Q+2H_0c(Q/(6H_0^2))^{n+1} + 2\Lambda)/G_N + bT$. \\
	
	We can recover $\Lambda$CDM Hubble function in two forms. The first form is to include the cosmological constant to account for the cosmological expansion. The second one is to choose an effective EoS that drives the acceleration and deceleration phases. However, $f(Q,T)$ theories do not require a cosmological constant to show accelerating and decelerating phases since they can be geometrical descriptions of dark energy as shown in \cite{arora2020f}.\\
	
	These $f(Q,T)$ theories need to be explored deeper, specifically the last model \ref{model:6} because it challenged $\Lambda$CDM at a background perspective. {A future study regarding the perturbation approach is crucial   to test whether our models constitute a strong alternative to the $\Lambda$CDM model which needs to be done by using Cosmic Microwave Background (CMB) and Large Scale Structure data (LSS)}.  Besides, gravitational waves are a great way to test modified gravity theories \cite{belgacem2019testing}. We can also use this data to test the $f(Q,T)$ theories, to enhance the constraints of the models and to quantify deviations from GR. 

\section*{Acknowledgments}
We are thankful to the referee for the detailed reading of our paper at all the stages of revision and for the time given to point out the issues that needed assessment. We are grateful to Dr. Celia Escamilla-Rivera for giving us some suggestions to improve the overall quality of this paper. Numerical computations for this research were done in the Tochtli cluster that is supported by the ICN-UNAM.

%\begin{figure}[pb]
    %\centering
    %\includegraphics[scale=0.60]{overlapping/model1.jpeg}
    %\includegraphics[scale=0.30]{overlapping/model2.jpeg}
    %\includegraphics[scale=0.40]{overlapping/model3.jpeg}
    %\includegraphics[scale=0.24]{overlapping/model4.jpeg}
    %\includegraphics[scale=0.30]{overlapping/model5.jpeg}
    %\caption{$\sigma$ and 2$\sigma$ confidence regions of the parameters $\Theta = \{ h, \Omega_\Lambda, $b$, $c$, $n$ \}$ with the CC, Pantheon and Pantheon+CC compilations. Top left: $\Lambda CDM$, top right: $f(Q,T) = -Q/G_N + bT$, middle left: $f(Q,T) = -(Q+2\Lambda)/G_N + bT$, middle right: $f(Q,T) = -(Q+2H_0^2 c (Q/(6H_0^2))^{n+1})/G_N + bT$, bottom:$f(Q,T) = -(Q+2 H_0 c (Q/(6H_0^2))^{n+1} + 2\Lambda)/G_N + bT$.}
   % \label{fig:trianglePlots}
%\end{figure}

\bibliographystyle{unsrt}
\bibliography{references}

\end{document}